\documentclass{article}
\usepackage{spconf,amsmath,graphicx}
\usepackage{IEEEtrantools}
\usepackage{cite}
\usepackage{amssymb,amsfonts}
\usepackage{algorithmic}
\usepackage{textcomp}
\usepackage{xcolor}
\usepackage{float}
\usepackage{amsthm}
\usepackage{graphicx}
\usepackage{epstopdf}
\usepackage{amsmath,bm}
\usepackage{amsfonts}
\usepackage{amssymb}
\usepackage{color}
\usepackage{multirow}
\usepackage{multicol}
\usepackage{soul,xcolor}

\definecolor{orange}{RGB}{0,112,192}
\title{LARGE-SCALE FADING PRECODING FOR MAXIMIZING THE PRODUCT OF SINRS}
\name{\"Ozlem Tugfe Demir and Emil Bj\"ornson \thanks{This work was partially supported by ELLIIT and the Wallenberg AI, Autonomous Systems and Software Program (WASP) funded by the Knut and Alice Wallenberg Foundation.}}
\address{{Department of Electrical Engineering (ISY), Link\"oping University, Sweden
	} \\
	{Email: \{ozlem.tugfe.demir, emil.bjornson\}@liu.se}}

\begin{document}
\ninept
\maketitle
\begin{abstract}
This paper considers the large-scale fading precoding design for mitigating the pilot contamination in the downlink of multi-cell massive MIMO (multiple-input multiple-output) systems. Rician fading with spatially correlated channels are considered where the line-of-sight (LOS) components of the channels are randomly phase-shifted in each coherence block. The large-scale fading precoding weights are designed based on maximizing the product of the signal-to-interference-plus-noise ratios (SINRs) of the users, which provides a good balance between max-min fairness and sum rate maximization. The spectral efficiency (SE) is derived based on the scaled least squares (LS) estimates of the channels, which only utilize the despreaded pilot signals without any matrix inversion. Simulation results show that the two-layer large-scale fading precoding improves the SE of almost all users compared to the conventional single-layer precoding. 
\end{abstract}
\vspace{-0.35cm}
\section{Introduction}
\vspace{-0.25cm}
\label{sec:intro}
Massive MIMO (multiple-input multiple-output) is one of the key technologies of 5G cellular systems and commercial deployments started in 2018 \cite{massive_mimo_reality}. Massive MIMO provides high spectral efficiency (SE) to a large number of users using the same time-frequency resources by an efficient spatial multiplexing and large number of antennas \cite{chien_lsfd,emil_book,erik_book, utshick, caire2, caire1}. In \cite{marzetta}, it is shown that the SE of the users is limited by pilot contamination which is the result of interference from the other cells' users whose pilot signals are not orthogonal to the considered cell's users. One trivial remedy is to increase the length of the pilot signals to make them all orthogonal throughout the network. However, due to limited coherence block length in practical communication systems, this approach is not efficient \cite{chien_lsfd},\cite{interference_marzetta}. Another solution to pilot contamination is to exploit the spatial correlation among the BS antennas \cite{unlimited}. This method alleviates the performance upper bound that was observed for uncorrelated channels in \cite{marzetta}, but, the precoding technique from \cite{unlimited} has high computational complexity.
\vspace{-0.4cm}
\subsection{Related Works}
\vspace{-0.2cm}
In \cite{interference_marzetta}, large-scale fading precoding (LSFP) and decoding (LSFD) approaches were elaborated, which require a limited cooperation between BSs and eliminate pilot contamination as the number of antennas goes to the infinity. This prior work considered spatially uncorrelated Rayleigh fading and the LSFP and LSFD weights are optimized using the max-min fairness criterion. Then, \cite{chien_lsfd} and \cite{uplink_lsfd} derived the SE with LSFD in the uplink for spatially correlated Rayleigh fading channels. There are no comparable studies for LSFP.

\subsection{Contributions}
\vspace{-0.15cm}
To the best of authors' knowledge, this paper is the first work that considers the spatially correlated Rician fading with random phase shifts for the design of LSFP. The main contributions are:
\begin{itemize}
	 
	\item We derive the SE for finite number of antennas for LSFP and least squares (LS) channel estimate-based local precoding. 
	
	\item We propose a successive convex approximation (SCA)-based  max product signal-to-interference-plus-noise ratio (SINR) optimization algorithm with guaranteed convergence. The variables are the transmit power and LSFP weights. 
	
	\item We show that LSFP improves the SE of the users uniformly compared to the single-layer precoding with cooperative power control based on the max product SINR.
\end{itemize}
\vspace{-0.3cm}
\section{System Model}
\vspace{-0.3cm}
We consider a cellular network with $L$ cells. Each cell is composed of an $M$-antenna base station (BS) and $K$ single-antenna users. We assume  all BSs are connected to a central network controller in accordance with the existing literature \cite{chien_lsfd}, \cite{interference_marzetta}. The conventional block-fading model \cite{erik_book} is assumed where the channel between each BS antenna and user is a complex static scalar in one coherence block of $\tau_c$ channel uses and take independent realization in each block. In this paper, we assume time division duplex (TDD) operation and, hence, channel reciprocity holds. We concentrate on the downlink part of the transmission where the BSs serve the users in their cells with the aid of LSFP at the central network controller. Each coherence block is divided into two phases: uplink training and downlink data transmission. In the uplink training phase, all users send their  pilot sequences which have length $\tau_p$ and the BSs estimate the channel information to design local precoding vectors that are matched to the estimated small-scale fading. The remaining $\tau_c-\tau_p$ samples are used for downlink data transmission. In accordance with the existing literature on massive MIMO, no downlink pilots are sent to the users and the users rely on channel statistics \cite{emil_book}.

Let ${\bf g}_{lk}^{r}\in \mathbb{C}^{M}$  denote the channel vector between user $k$ in cell $l$ and BS $r$.  We consider spatially correlated Rician fading channels, which is the first novelty of this paper in the context of LSFP. This means each channel realization can be expressed as
\vspace{-0.15cm}
\begin{align} \label{eq:channel}
& {\bf g}_{lk}^{r}=e^{j\theta_{lk}^r}{{\bf \bar{g}}}_{lk}^{r}+{\bf \tilde{g}}_{lk}^{r},
\end{align}
where $e^{j\theta_{lk}^r}{\bf \bar{g}}_{lk}^{r}\in \mathbb{C}^{M}$  denotes the line-of-sight (LOS) component with some phase shift $\theta_{lk}^r$ common to all the antennas. The other term of the channel, i.e., ${\bf \tilde{g}}_{lk}^{r}$, is the non-line-of-sight (NLOS) component and it is circularly symmetric Gaussian random vector with spatial covariance matrix ${\bf R}_{lk}^{r} \in \mathbb{C}^{M\times M}$, i.e., ${\bf \tilde{g}}_{lk}^{r} \sim \mathcal{N}_{\mathbb{C}}({\bf 0}_M,{\bf R}_{lk}^{r})$. Note that the vectors $\{{\bf \bar{g}}_{lk}^{r}\}$ and covariance matrices $\{{\bf R}_{lk}^{r}\}$ describe the long-term channel effects and change more slowly compared to small-scale fading characteristics. We assume that all BSs have the knowledge of $\{{\bf \bar{g}}_{lk}^{r}\}$ and covariance matrices ${\bf R}_{lk}^r$ in accordance with the multi-cell massive MIMO literature \cite{chien_lsfd}, \cite{ozge_massive}. However compared to most of the existing literature, we consider a more realistic scenario in which phase of the LOS component varies at the same pace as the small-scale fading, due movement, and the phase is unknown. We assume the random phase shifts $\{\theta_{lk}^r\}$ are distributed uniformly on $[0,2\pi)$ \cite{ozge_cell_free}.

\vspace{-0.15cm}
\subsection{Uplink Pilot Transmission}
\vspace{-0.15cm}
All the cells use a common set of $\tau_p=K$ mutually orthogonal pilots where the pilots are distributed among the $K$ users in each cell in a disjoint manner. Let $\bm{\varphi}_k \in \mathbb{C}^{\tau_p}$ denote the pilot sequence which is assigned to the user $k$ in each cell where $||\bm{\varphi}_{k}||^2=\tau_p$ and $\bm{\varphi}_k^H\bm{\varphi}_{k^{\prime}}=0$, $\forall k^{\prime}\neq k$. Due to the pilot re-use between different cells, there is interference in the pilot transmission and so-called pilot contamination occurs.

During the uplink training phase, the received pilot signal ${\bf Z}_l\in \mathbb{C}^{M\times \tau_p}$ at BS $l$ is given by
\vspace{-0.15cm}
\begin{align}\label{eq:pilot}
& {\bf Z}_{l}=\sum_{r=1}^L\sum_{k=1}^{K}\sqrt{\eta}{\bf g}_{rk}^{l}\bm{\varphi}_{k}^T+{\bf N}_l,
\end{align}
where $\eta$ is the pilot transmit power of the UEs and the additive noise matrix ${\bf N}_{l}\in \mathbb{C}^{M \times \tau_p}$ has i.i.d. $\mathcal{N}_{\mathbb{C}}(0,\sigma^2)$ random variables. Then, the sufficient statistics for the channel information of user $k$ in cell $l$ is obtained as 
\vspace{-0.15cm}
\begin{align}\label{eq:suff-stats} 
& {\bf z}_{lk}=\frac{{\bf Z}_l\bm{\varphi}_k^{*}}{\sqrt{\tau_p}}=\sqrt{\tau_p\eta}\sum_{r=1}^L {\bf g}_{rk}^l+{\bf \tilde{n}}_{lk},
\end{align}
where $\tilde{\bf n}_{lk}\triangleq{\bf N}_l\bm{\varphi}_k^{*}/\sqrt{\tau_p}$ has the distribution $\mathcal{N}_{\mathbb{C}}({\bf 0}_M,\sigma^2{\bf I}_M)$. 

If the phase shifts are not known, deriving an MMSE-based channel estimator is very hard since we do not have a linear Gaussian signal model. Instead, we can use LS-based channel estimate. Note that ${\bf z}_{lk}$ is the scaled version of LS channel estimate for the users that share the pilot sequence $k$. The covariance matrix of zero-mean non-Gaussian ${\bf z}_{lk}$ is given by
\vspace{-0.15cm}
\begin{align}
& {\bf \Psi}_{lk}\triangleq\mathbb{E}\{{\bf z}_{lk}{\bf z}_{lk}^H\}= \tau_p\eta\sum_{r=1}^L\left({\bf R}_{rk}^l+{\bf \bar{g}}_{rk}^l({\bf \bar{g}}_{rk}^l)^H\right)+\sigma^2{\bf I}_M \label{eq:Psibar}.
\end{align}

\vspace{-0.25cm}
\subsection{Downlink Data Transmission}
\vspace{-0.15cm}
Let $s_{lk}$ denote the zero-mean unit-variance downlink symbol for the user $k$ in the cell $l$. In accordance with the previous work \cite{interference_marzetta}, we assume all the downlink symbols are accessible to the central network controller for LSFP. In the first step, the central network controller computes the symbols to be transmitted from each BS by conducting large-scale fading precoding using the long-term statistics of the channel. Let $(a_{rk}^l)^{*}$ denote the complex weight applied to the data symbol of user $k$ in cell $r$ for the transmission of the combined signal $v_{lk}$ from the BS $l$, i.e.,
\vspace{-0.15cm}
\begin{align}\label{eq:vlk} 
& v_{lk} =\sum_{r=1}^L(a_{rk}^{l})^{*}s_{rk}.
\end{align} 
Here, $v_{lk}$ is the precoded signal to be transmitted from BS $l$ for the users with index $k$. Hence, LSFP is applied to the data symbols of users sharing the same pilot sequence and each BS transmits a linear combination of pilot-sharing users' signals in an effort to precancel the pilot-contaminated interference that occur between them.

In the second step, the central network controller sends the precoded symbols $\{v_{lk}\}_{k=1}^K$ to the BS $l$. Note that the power control is applied through the LSFP coefficients $\{a_{rk}^l\}$. Hence, there is no need for an additional localized power control at the BSs. In the last step, the BS $l$ conducts local precoding based on the channel information obtained from the uplink pilot transmission.  Using ${\bf z}_{lk}^*$ as the precoding vector for the users who share the pilot sequence $k$, the transmitted signal from the BS $l$ is 
\vspace{-0.15cm}
\begin{align}\label{eq:xl} 
& {\bf x}_l =\sum_{k=1}^K{\bf z}_{lk}^{*}v_{lk}.
\end{align}
\vspace{-0.5cm} 
\section{Downlink SE Analysis}
\vspace{-0.3cm} 
In this section, we will derive a closed-form SE expression when using the considered local precoding based on the sufficient statistics. The use-and-then-forget capacity bounding technique, which is common in massive MIMO literature, will be utilized in order to obtain closed-form lower bounds on the downlink ergodic capacity \cite{emil_book}.

The received signal at user $k$ in the cell $l$ is 
\vspace{-0.15cm}
\begin{align} \label{eq:ylk}
& y_{lk}=\sum_{r=1}^L({\bf g}_{lk}^{r})^T{\bf x}_{r}+ n_{lk},
\end{align}
where $n_{lk}$ is the additive Gaussian noise at the receiver of user $k$ in cell $l$ with zero-mean and variance $\sigma^2$. Let us rewrite \eqref{eq:ylk} as
\vspace{-0.15cm}
\begin{align} \label{eq:ylk2} 
& y_{lk}=\text{DS}_{lk}s_{lk}+\text{BU}_{lk}s_{lk}+\sum_{\substack{r=1\\r\neq l}}^L\text{PC}_{lk}^{r}s_{rk} \nonumber \\
&\hspace{1cm}+\sum_{r=1}^L\sum_{\substack{k^{\prime}=1\\ k^{\prime}\neq k}}^K\text{NI}_{lk}^{rk^{\prime}}s_{rk^{\prime}}+n_{lk},
\end{align}
where $\text{DS}_{lk}$, $\text{BU}_{lk}$, $\text{PC}_{lk}^{r}$, and $\text{NI}_{lk}^{rk^{\prime}}$ represent the strength of the desired signal, the beamforming gain uncertainty, the pilot contamination, and the non-coherent interference which are defined as
\vspace{-0.15cm}
\begin{align} 
&\text{DS}_{lk}=\sum_{r=1}^L(a_{lk}^{r})^{*}\mathbb{E}\{{\bf z}_{rk}^H{\bf g}_{lk}^{r}\} \label{eq:DS}, \\
&\text{BU}_{lk}=\sum_{r=1}^L(a_{lk}^{r})^{*}\Big({\bf z}_{rk}^H{\bf g}_{lk}^{r}-\mathbb{E}\{{\bf z}_{rk}^H{\bf g}_{lk}^{r}\}\Big) \label{eq:BU}, \\
& \text{PC}_{lk}^{r}=\sum_{n=1}^L(a_{rk}^{n})^{*}{\bf z}_{nk}^H{\bf g}_{lk}^{n}, \label{eq:PC} \\
&\text{NI}_{lk}^{rk^{\prime}}= \sum_{n=1}^L(a_{rk^{\prime}}^{n})^{*}{\bf z}_{nk^{\prime}}^H{\bf g}_{lk}^{n} \label{eq:NI}.
\end{align}
Note that the users do not know the actual value of the effective channel components ${\bf z}_{rk}^H{\bf g}_{lk}^{r}$ for the desired symbol, but only the expected value of the effective channel, which is $\text{DS}_{lk}$ in \eqref{eq:DS}. The expected value is close to the effective channel in massive MIMO, if the precoding is selected based on the channel vector. Hence, the beamforming gain uncertainty resulting from imperfect channel state information in \eqref{eq:BU} is treated as interference. The pilot contamination from the users that use the same pilots in other cells and the non-coherent interference from the remaining users are the other sources of interference. The following lemma presents a lower bound on the downlink ergodic capacity by treating the signals other than desired signal as additive white Gaussian noise and using the use-and-then-forget capacity bounding technique \cite[Sec. 4]{emil_book}.

{\bf Lemma 1:}  The downlink ergodic capacity of the user $k$ in cell $l$ for the given LSFP weights, is lower bounded by 
\begin{align}\label{eq:capacity}
R_{lk}= \left(1-\frac{\tau_p}{\tau_c}\right)\log_2\left(1+\text{SINR}_{lk}\right),
\end{align}
where $\text{SINR}_{lk}$ is the effective SINR for user $k$ in cell $l$ and given by
\begin{align}\label{eq:sinr}
&\text{SINR}_{lk}=\nonumber \\
&\frac{|\text{DS}_{lk}|^2}{\mathbb{E}\{|\text{BU}_{lk}|^2\}+\sum\limits_{\substack{r=1 \\ r\neq l}}^L\mathbb{E}\{|\text{PC}_{lk}^{r}|^2\}+\sum\limits_{\substack{r=1}}^L\sum\limits_{\substack{k^{\prime}=1 \\ k^{\prime}\neq k}}^K\mathbb{E}\{|\text{NI}_{lk}^{rk^{\prime}}|^2\}+\sigma^2}.
\end{align}
\begin{IEEEproof}
 The lower bound expression in \eqref{eq:capacity} follows from \cite{erik_book} by noting that the interference  terms are mutually uncorrelated. 
 \end{IEEEproof}

Let us define the following vectors and matrices for ease of notation in the following parts of the paper:
\begin{align}
& \bm{a}_{lk}\triangleq [ \ a_{lk}^1 \ \ldots \ a_{lk}^{L} \ ]^T \in \mathbb{C}^{L},  \label{eq:alk}  \\
& \bm{b}_{lk}\triangleq [ \ b_{lk}^1 \ \ldots \ b_{lk}^{L}  \ ]^T \in \mathbb{C}^{L}, \ \ \ b_{lk}^r\triangleq \mathbb{E}\{{\bf z}_{rk}^H{\bf g}_{lk}^{r}\} \label{eq:blk},  \\
&\bm{C}_{lkk^{\prime}}\in \mathbb{C}^{L \times L}, \ \ \ c_{lkk^{\prime}}^{rn}\triangleq \mathbb{E}\{{\bf z}_{rk^{\prime}}^H{\bf g}_{lk}^{r}({\bf g}_{lk}^{n})^H{\bf z}_{nk^{\prime}}\} \label{eq:Clk},
\end{align}
where $c_{lkk^{\prime}}^{rn}=\big[\bm{C}_{lkk^{\prime}}\big]_{rn}$ is the $(r,n)$th element of the matrix $\bm{C}_{lkk^{\prime}}$.
Using the above definitions, the terms in the SINR expression in \eqref{eq:sinr} can be expressed as
\begin{align} 
&|\text{DS}_{lk}|^2=|\bm{a}_{lk}^H\bm{b}_{lk}|^2, \label{eq:DS2} \\
&\mathbb{E}\{|\text{BU}_{lk}|^2\}=\bm{a}_{lk}^H\bm{C}_{lkk}\bm{a}_{lk}-|\bm{a}_{lk}^H\bm{b}_{lk}|^2, \label{eq:BU2} \\
& \mathbb{E}\{|\text{PC}_{lk}^{r}|^2\}=\bm{a}_{rk}^H\bm{C}_{lkk}\bm{a}_{rk}, \label{eq:PC2} \\
&\mathbb{E}\{|\text{NI}_{lk}^{rk^{\prime}}|^2\}=\bm{a}_{rk^{\prime}}^H\bm{C}_{lkk^{\prime}}\bm{a}_{rk^{\prime}} \label{eq:NI2}.
\end{align}
The following theorem present the lower bound on the downlink ergodic capacity for the local precoding based on $\{{\bf z}_{lk}\}$ in \eqref{eq:suff-stats}.

{\bf Theorem 1:} For a given set of LSFP coefficients, the SE of user $k$ in cell $l$ for the local precoding vectors $\{{\bf z}_{lk}^{*}\}$ is given in \eqref{eq:capacity} with $\text{SINR}_{lk}$ as
\begin{align}\label{eq:SE1}
&\text{SINR}_{lk}=\nonumber \\
&\frac{|\bm{a}_{lk}^H\bm{b}_{lk}|^2}{\sum\limits_{r=1}^L\bm{a}_{rk}^H\bm{C}_{lkk}\bm{a}_{rk}-|\bm{a}_{lk}^H\bm{b}_{lk}|^2+\sum\limits_{\substack{r=1}}^L\sum\limits_{\substack{k^{\prime}=1 \\ k^{\prime}\neq k}}^K\bm{a}_{rk^{\prime}}^H\bm{C}_{lkk^{\prime}}\bm{a}_{rk^{\prime}}+\sigma^2},
\end{align}
where the elements of $\bm{b}_{lk}$ and $\bm{C}_{lkk^{\prime}}$ are given by
\begin{align}
b_{lk}^{r}&=\sqrt{\tau_p\eta}\left(\left({\bf \bar{g}}_{lk}^r\right)^H{\bf \bar{g}}_{lk}^r+\text{tr}\left({\bf R}_{lk}^{r}\right)\right), \label{eq:blk1} \\
c_{lkk}^{rr}&=\tau_p\eta\left(\text{tr}\left({\bf R}_{lk}^{r}\right)\right)^2+2\tau_p\eta({\bf \bar{g}}_{lk}^r)^H{\bf \bar{g}}_{lk}^r\text{tr}\left({\bf R}_{lk}^{r}\right)\nonumber \\
&+\text{tr}\left({\bf \Psi}_{rk}\left({\bf R}_{lk}^{r}+{\bf \bar{g}}_{lk}^r({\bf \bar{g}}_{lk}^r)^H\right)\right) \label{eq:clkk-rrb}, \\
c_{lkk}^{rn}&=b_{lk}^r(b_{lk}^n)^{*}, \ \ \ r\neq n, \label{eq:clkk-rn-b} \\
c_{lkk^{\prime}}^{rr}&=\text{tr}\left({\bf \Psi}_{rk^{\prime}}\left({\bf R}_{lk}^{r}+{\bf \bar{g}}_{lk}^r({\bf \bar{g}}_{lk}^r)^H\right)\right), \ \  k^{\prime} \neq k, \label{eq:clkk2-rr-b} \\
c_{lkk^{\prime}}^{rn}&=0, \ \ \ k^{\prime} \neq k, \ r \neq n. \label{eq:clkk2-rn-b}
\end{align}
\begin{IEEEproof} The proof follows from the calculating expectations using the statistics of Rician fading channels with phase shifts in the LOS components. The details are omitted due to space limitation.
\end{IEEEproof}

\vspace{-0.2cm}
\section{LSFP That Maximizes the Product SINR}

Note that the LSFD weights in the uplink can be selected independently for each user, as shown in \cite{chien_lsfd}. However, for the LSFP, a joint design is needed since it affects the interference level of all users. In this paper, we design the LSFP weights $\{a_{rk}^l\}$ based on the proportional fairness criterion \cite{emil_book}, \cite{luca_max_product},\cite{amin}, which aims to maximize the product of all SINRs under individual BS transmit power constraints. The max product SINR is an optimization objective that obtains a good balance between max-min fairness, which limits the network-wide performance by focusing on the users with the worst channel conditions \cite{amin}, and sum rate maximization, which does not guarantee any user fairness. Maximizing the product of SINRs, which are given in \eqref{eq:SE1}, corresponds to a modified version of sum rate maximization where more emphasis is put on the weakest users by excluding 1 in the logarithm. Please note that, the method we propose can also be used for sum rate maximization with a slight modification in the objective.

The long-term transmit power of BS $l$ is given by
\begin{align}
P_{l}=\mathbb{E}\left\{{\bf x}_l^H{\bf x}_l\right\}=\sum_{k=1}^K\text{tr}\left({\bf \Psi}_{lk}\right)\sum_{r=1}^L|a_{rk}^l|^2.
\end{align}
Using this notation, the max product SINR problem can be expressed as 
\begin{align}
&\underset{\left\{\bm{a}_{lk}\right\}}{\text{maximize}} \ \ \ \prod_{l=1}^L\prod_{k=1}^K \text{SINR}_{lk} \label{eq:obj} \\
&\text{subject to} \ \ \ P_{l}\leq\rho_d, \ \ l=1,\ldots,L,  \label{eq:const}
\end{align}
where $\rho_d$ is the maximum downlink transmission power of each BS.

Note that the problem in (\ref{eq:obj})-(\ref{eq:const}) is not convex. In fact, the objective function in \eqref{eq:obj} is a concave function of the SINRs and the per-BS power constraints in \eqref{eq:const} are convex. However, SINRs are not linear functions of the LSFP vectors. To solve the non-convexity problem, we will develop a successive convex approximation (SCA) method to solve the problem in an iterative manner where a convex problem is solved at each iteration. Since, the objective function is upper bounded and will be improved at each iteration, this iterative scheme is guaranteed to converge.

{\bf Remark:} The problems in \cite{emil_book}, \cite{amin} for maximizing the product of SINRs can be solved optimally by casting them in a geometric programming form since they consider single-layer precoding and decoding. However, the problem given in (\ref{eq:obj})-(\ref{eq:const}) has a different structure and does not admit a geometric programming formulation. Furthermore, the per-BS power constraints imply that uplink/downlink duality cannot be utilized to simplify the problem structure.

Let us reformulate the problem (\ref{eq:obj}-\ref{eq:const}) as follows:
\begin{align}
&\underset{\left\{\bm{a}_{lk}, \ t_{lk}, \ u_{lk} \right\}}{\text{maximize}} \ \ \ \prod_{l=1}^L\prod_{k=1}^K t_{lk} \label{eq:objb} \\
&\text{subject to} \nonumber \\
& \hspace{0.2cm} \sum_{r=1}^L\sum_{k^{\prime}=1}^K\bm{a}_{rk^{\prime}}^H{\bf C}_{lkk^{\prime}}\bm{a}_{rk^{\prime}}-\bm{a}_{lk}^H\bm{b}_{lk}\bm{b}_{lk}^H\bm{a}_{lk}+\sigma^2\leq u_{lk}, \label{eq:const1} \\
&\hspace{0.2cm} \left(\Re\{\bm{a}_{lk}^H\bm{b}_{lk}\}\right)^2\geq t_{lk}u_{lk}, \ \  l=1,\ldots,L, \  k=1,\ldots,K, \label{eq:const2} \\
&\hspace{0.2cm} P_{l}\leq\rho_d, \ \ l=1,\ldots,L, \label{eq:const3} 
\end{align}
where we utilize that $\bm{a}_{lk}$ can have any phase in \eqref{eq:const2} and therefore we can rotate it to get a real-valued number. Note that all the constraints are convex except \eqref{eq:const2} and the objective function is concave (note that ${\bf C}_{lkk}-\bm{b}_{lk}\bm{b}_{lk}^H$ is positive semidefinite.) Using the upper bound $\left(0.5u_{lk}^{(j-1)}/t_{lk}^{(j-1)}\right)t_{lk}^2+\left(0.5t_{lk}^{(j-1)}/u_{lk}^{(j-1)}\right)u_{lk}^2\geq t_{lk}u_{lk}$ as in \cite{valid_surrogate},\cite{ozlem_swipt}, we can approximate the problem in (\ref{eq:objb})-(\ref{eq:const3}) as a convex problem where a feasible solution to the original problem is obtained at each iteration if the convex approximation is feasible. Here, $(.)^{(j-1)}$ denotes the value of an optimization variable in iteration $j-1$. To guarantee the feasibility at the initialization of the algorithm, we include some positive variables at the right side of \eqref{eq:const1} and these are enforced to become zero at the later iterations by including a corresponding penalty with a large weight in the objective function. This idea is used in feasible point pursuit-successive convex approximation (FPP-SCA) method in \cite{fpp}. At the $j^{\textrm{th}}$ iteration, the following problem is solved:
\begin{align}
&\underset{\left\{\bm{a}_{lk}, \ t_{lk}, \ u_{lk}, \ f_{lk} \right\}}{\text{maximize}} \ \ \ \prod_{l=1}^L\prod_{k=1}^K t_{lk}-\lambda\sum_{l=1}^L\sum_{k=1}^Kf_{lk} \label{eq:objbb} \\
&\text{subject to} \ \ \ \sum_{r=1}^L\sum_{k^{\prime}=1}^K\bm{a}_{rk^{\prime}}^H{\bf C}_{lkk^{\prime}}\bm{a}_{rk^{\prime}}-\bm{a}_{lk}^H\bm{b}_{lk}\bm{b}_{lk}^H\bm{a}_{lk}+\sigma^2 \nonumber \\
&\hspace{2cm}\leq u_{lk}+f_{lk},\label{eq:const1b} \\
&\hspace{1.2cm} \ \ \ \left(\Re\{\bm{a}_{lk}^H\bm{b}_{lk}\}\right)^2\geq \frac{u_{lk}^{(j-1)}}{2t_{lk}^{(j-1)}}t_{lk}^2+\frac{t_{lk}^{(j-1)}}{2u_{lk}^{(j-1)}}u_{lk}^2, \label{eq:const2b} \\
& \hspace{1.2cm} \ \ \ f_{lk}\geq 0, \ \  l=1,\ldots,L, \ \ k=1,\ldots,K,  \\
&\hspace{1.2cm} \ \ \ P_{l}\leq\rho_d, \ \ l=1,\ldots,L, \label{eq:const3b} 
\end{align}
where $\{f_{lk}\geq0\}$ are the auxiliary variables to make all the subproblems feasible. When the problem at the $j^{\textrm{th}}$ iteration is feasible and a large positive weight $\lambda$ is used, these slack variables tend to go to zero, and we obtain the original subproblem \cite{fpp}. Note that (\ref{eq:objbb}-\ref{eq:const3b}) is convex since \eqref{eq:const2b} is a second-order cone constraint.

	\vspace{-0.3cm}
\section{Numerical Results}
	\vspace{-0.3cm}
We compare the SE of the optimized LSFP scheme (two-layer decoding) with standard single-layer (local) precoding. We consider two power-control schemes for single-layer precoding. The first one is non-cooperative  and  the downlink signal power of user $k$ in cell $l$ at its serving BS $l$ is proportional to $\sqrt{\text{tr}({\bf \Psi}_{lk})}$ and total transmitted power from each BS is $\rho_d$ following the power-control approach in \cite{giovanni}. In the figures, this scheme is denoted by LPC (local power control). The second power control is more advanced and we optimize the powers for proportional fairness as described in this paper but under the additional constraint that only the $l^{\textrm{th}}$ entry of $\bm{a}_{lk}$ is non-zero, which means that each BS only transmit data to their own users. This scheme is denoted by CPC (cooperative power control) in the figures. Hence, we optimize both the two-layer precoding and single-layer precoding with CPC using max product SINR criteria for a fair comparison. We consider the Rician fading multi-cell setup in \cite{ozge_massive}, except that in our scenario, the phases of the LOS components are shifted randomly in every coherence block. There are $L=4$ cells in the network where each cell occupies a 150\,m$\times$150\,m square area with the BS at the center. The number of antennas at each BS is $M=200$. The uplink pilot power is $\eta=0.05$ W and the maximum downlink transmit power is $\rho_d=2$ W. The noise variance is $\sigma^2=-96$ dBm. We present the results of 100 different setups where the users are dropped in the cells uniformly with at least 20 m distance to the BSs in accordance with the urban microcell model in \cite{channel}. 200 channel realizations are considered for each setup for the Monte-Carlo verification of the derived SEs.

In Fig.~1, we plot the average sum SE per cell versus the number of users per cell which is $K$. All the analytical results are verified by Monte Carlo estimations which are shown by red dots. We notice that for all $K$, two-layer precoding improves the sum rate substantially. The gain compared to the single-layer precoding with CPC changes  between 15\% and 32\%, which is higher than the relative performance improvements for the uplink counterpart in \cite{chien_lsfd} with maximum ratio processing. In comparison to the LPC scheme, two-layer precoding provides between 26\% and 48\% gain in sum SE.
\vspace{-0.2cm}
\begin{figure}[t!]
	\includegraphics[trim={1cm 0cm 0.8cm 0.5cm},clip,width=3.4in]{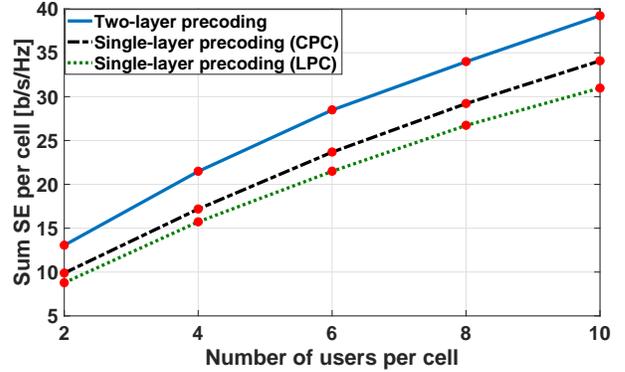}
	\vspace{-0.8cm}
	\caption{Average sum SE per cell for different $K$ values.}
	\label{fig:fig1}
	\vspace{-0.3cm}
\end{figure}

In order to see the impact of the product SINR maximization on the individual user SEs, we consider the cumulative distribution function (CDF) of the SE per user for $K=6$ in Fig.~2. It is seen that the LSFP provides higher SE consistently for almost all the users. The median SE is increased by around 50\% and 31\% by two-layer precoding compared to the local and cooperative power control-based single layer precoding, respectively.

\vspace{-0.25cm}

\begin{figure}[t!]
	\includegraphics[trim={1cm 0cm 0.8cm 0.5cm},clip,width=3.4in]{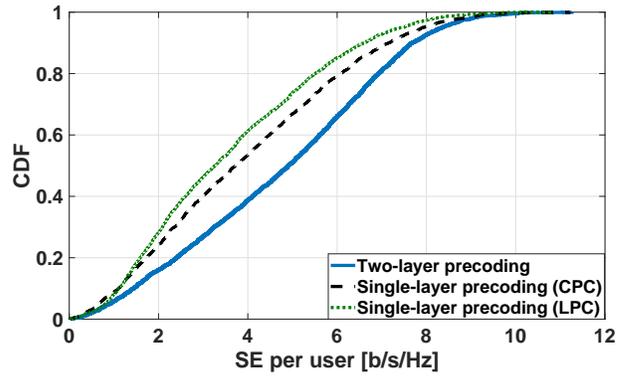}
	\vspace{-0.8cm}
	\caption{CDF of the SE per user for $K=6$.}
	\label{fig:fig2}
	\vspace{-0.3cm}
\end{figure}
\vspace{-0.1cm}
\section{Conclusion}
\vspace{-0.2cm}
In this paper, we consider LSFP design to mitigate the adverse effects of pilot contamination in a realistic Rician fading environment where the LOS components of the channels  are corrupted  by random phase shifts which are not known at the BSs. We derive the closed-form SEs for a simple local precoding scheme. We propose an SCA-based algorithm for the optimization of LSFP weights at the central network controller to maximize the product of the SINRs. Simulation results verify the effectiveness of the max product SINR algorithm for both single and two-layer precoding. Furthermore, both the individual SE and sum SEs are improved by the two-layer precoding employing LSFP in comparison to standard single-layer precoding.

\end{document}